\documentclass[aps,prl,twocolumn,superscriptaddress,floatfix]{revtex4-2}
\usepackage[dvips]{color}
\usepackage{graphicx}
\usepackage{amssymb}
\usepackage{amsmath}
\usepackage{amsfonts}
\usepackage{bbold}
\usepackage{soul}
\usepackage{xcolor}

\begin{document}

\title{Identifying primes from entanglement dynamics}

\author{A. L. M. Southier}
\affiliation{Departamento de F\'{\i}sica, Universidade Federal do Paran\'a, 81531-990, Curitiba, PR, Brazil}\author{L. F. Santos}
\affiliation{Department of Physics, University of Connecticut, Storrs, CT, USA}
\author{P. H. Souto Ribeiro}
\affiliation{Departamento de F\'{\i}sica, Universidade Federal de Santa Catarina, 88040-900, Florian\'{o}polis, SC, Brazil}
\author{A. D. Ribeiro}
\affiliation{Departamento de F\'{\i}sica, Universidade Federal do Paran\'a, 81531-990, Curitiba, PR, Brazil}

\begin{abstract}
Factorization is the most fundamental way to determine if a number $n$ is prime or composite. Yet, this approach becomes impracticable when considering large values of $n$, a difficulty that is exploited by cryptographic protocols. We propose an alternative method to decide the primality of a natural number, that is based on the analysis of the evolution of the linear entanglement entropy. Specifically, we show that a singular behavior in the amplitudes of the Fourier series of this entropy is associated with prime numbers. We also discuss how this idea could be experimentally implemented and examine possible connections between our results and the zeros of the Riemann zeta function. 
\end{abstract}

\maketitle

Prime numbers have captured the attention of researchers for centuries. In pure mathematics, given the prominent role of primes in factorizing a positive integer $n$, much effort has been made to unravel patterns in their distribution over the set of natural numbers $\mathbb{N}$~\cite{ribenboim}. Remarkable results in this direction involve the counting function $\pi(n)$, which gives the number of primes less than or equal to $n$. The complete knowledge of $\pi(n)$ would imply being able to determine the position of each prime number $p$, since jumps of this function expose their presence. However, achieving the counting function with satisfactory accuracy for large values of $n$ has been shown to be practically impossible. In applied mathematics, this lack of knowledge is exploited for cryptographic protocols~\cite{koblitz}, such as the RSA  (Rivest-Shamir-Adleman) algorithm. To break the RSA protocol, one needs to find the prime factors of a huge $n$, which would require the implementation of Shor's algorithm in a quantum computer.

Another fascinating result involving primes is their connection with the zeros of the Riemann zeta function~$\zeta(s) $, where $s$ is a complex variable, defined as~\cite{edwards}
\begin{equation}
\zeta(s) \equiv \sum_{n=1}^{\infty}\frac{1}{n^s} = 
\prod_{p} \frac{1}{1-p^{-s}}=
\frac{\Gamma(1-s)}{2\pi i}\int_\gamma 
\frac{(-x)^s}{\mbox{e}^x-1} \frac{\mbox{d}x}{x}.
\label{RS}
\end{equation}  
The summation and the product above converge when the real part of the variable $s$ satisfies $\Re[s]>1$. The first equality in Eq.~(\ref{RS}) is due to Euler and makes evident the relation between $p$ and $\zeta(s)$.  In 1859~\cite{riemann}, Riemann achieved the expression shown on the right side of Eq.~(\ref{RS}), which involves a line integral along a particular path $\gamma$~\cite{path} and the Gamma function $\Gamma(1-s)\equiv\int_0^\infty x^{-s} \mbox{e}^{-x}\mbox{d}x$. The expression is analytic for all values of $s$, except for a simple pole at $s=1$. In the region $\Re[s]>1$, the term with the integral recovers the other two expressions in Eq.~(\ref{RS}), so one considers it as their analytic continuation.  Using these ideas, Riemann~\cite{riemann} demonstrated how the complex zeros $s_0$ of $\zeta(s)$ encode the distribution of $p$. Specifically, he showed that a certain convergent series, running over all $s_0$, recovers the function $\pi(n)$. But for large values of $n$, the number of zeros required to accurately get the counting function is so large that the use of the series to identify $p$ becomes intractable. This sophisticated connection between $\zeta(s)$ and primes is but one of the  impressive results of Ref.~\cite{riemann}. Arguments in that work also gave origin to the Riemann hypothesis, which conjectures that $\Re[s_0]=\frac12$ for all nontrivial zeros of~$\zeta(s)$.

The almost mystical status of prime numbers has also reached the physics community, especially researchers working with quantum mechanics~\cite{hutchinson2011,wolf2020}. A particularly inspiring idea is the Hilbert-P\'olya conjecture that could lead to the proof of the Riemann hypothesis. The conjecture proposes that the nontrivial zeros of the Riemann zeta function, which supposedly fall on the critical line $\Re[s_0]=\frac12$, correspond to the eigenvalues of a Hermitian Hamiltonian operator. The first substantial evidence supporting this conjecture arose in the semiclassical studies carried out by Berry and Keating~\cite{keating99}. They compared the distribution of $s_0$ over the critical line with the equivalent function for the eigenvalues of a given Hamiltonian achieved through the Gutzwiller trace formula. In this way, they were able to identify a Hamiltonian operator that fulfills the conjecture. The operator is obtained by quantizing the classical Hamiltonian $H_{\mathrm cl} = xp$, where $x$ and $p$ are the particle position and momentum, respectively.

Inspired by the Hilbert-P\'olya conjecture and the Berry and Keating $xp$-model, several works~\cite{sierra2008,hutchinson2008,sierra2011,srednicki2011,sierra2012,wolf2014,muller2017} have aimed at interpreting, extending and circumventing technical difficulties of Ref.~\cite{keating99}. Other contributions looked for alternative physical systems, where the properties of $\zeta(s)$ could be identified. This has been done using quantum graph theory~\cite{richter2014}, many-body systems~\cite{mussardo2020}, wave-packet dynamics~\cite{schleich2010}, statistical mechanics of random energy landscapes~\cite{keating2012}, random matrix theory~\cite{keating1995,leboeuf2003}, quantum entanglement~\cite{schleich2013}, and there is also an experiment based on a periodically driven single qubit~\cite{guang2020}. In addition to these works, where the Riemann zeta function is the protagonist to link prime numbers and quantum physics, other quantum approaches deal directly with primes, studying, for instance, number factorization~\cite{schleich2018, martin2018} and the properties of prime states \cite{sierra2015,sierra2020}, which are superpositions of states corresponding to prime numbers.

In the present paper, we show that the dynamics of the linear entanglement entropy encodes prime and semiprime numbers.  Specifically, we demonstrate that the Fourier modes $c_n$ of this evolving function have amplitudes that present a singular behavior when $n$ corresponds to a prime, and we find curves $c_n^{f}$ for the location of families $f$ of semiprimes. This is shown for a system of two coupled harmonic oscillators and a system of two coupled spins. We discuss a possible experimental realization of this idea and speculate on how to link $c_n$ and the counting function~$\pi(n)$.

%%%%%%%%%%%%%%%%%%%%%%%%%%%%%%%%%%%%%%%%%%%%%
{\em Entanglement dynamics.---} We consider a system consisting of two interacting parts, $A$ and $B$, to which we assign distinct Hilbert spaces, $\mathcal{H}_A$ and $\mathcal{H}_B$, respectively.  The system is isolated and prepared in a pure state,  $\rho (0) = |\Psi(0)\rangle \langle \Psi(0) |$, which evolves according to the total Hamiltonian
\begin{equation}
H= H_A\otimes \mathbb{1}_B + \mathbb{1}_A \otimes H_B + \lambda H_{A} \otimes H_{B},
\label{Hoperator}
\end{equation}
where $\lambda$ is the coupling strength between the two subsystems. The linear entanglement entropy,
\begin{equation}
S_{L}(t) = 1 - \mathrm{Tr}
\left[ \rho_{A}^2(t) \right],
\label{Slin}
\end{equation}
where $\rho_{A} (t) \equiv\mathrm{Tr}_{B} [\rho (t)]$ is the reduced density matrix of system $A$, measures the bipartite entanglement in time between  $A$ and  $B$. When $\rho(t)$ is separable, $S_L (t)\!=\!0$, otherwise $0\!<\!S_{L}(t)\!<\!1$.

%%%%%%%%%%%%%%%%%%%%%%%%%%%%%%%%%%%%%%%%%%%%%
{\em Coupled oscillators.---} The first Hamiltonian that we consider describes two coupled harmonic oscillators, where $H_{A}$ is the Hamiltonian for part~$A$ with eigenvalues $\hbar\omega_0(n_{A}+\frac12)$ and eigenstates $|n_{A}\rangle$, and equivalently for part $B$. The initial state is the product of two canonical coherent states~\cite{gazeau},
\begin{equation}
|\Psi (0)\rangle =
|\alpha_{A}\,\alpha_{B} \rangle \equiv \mbox{e}^{-u/2} \!\!\!\!\!\!
\sum_{n_{A},n_{B}=0}^\infty 
\frac{\alpha_{A}^{n_A}}{\sqrt{n_{A}!}} 
\frac{\alpha_{B}^{n_{B}}}{\sqrt{n_{B}!}}
|n_{A}\,n_{B}\rangle ,
\nonumber
\end{equation}
where we chose $|\alpha_{A}|^2=|\alpha_{B}|^2 = \frac{u}{2}$.  Defining $\omega \equiv \hbar \lambda \omega_0^2$,
\begin{equation}
S_L^{osc}(t) = 
1 - \mbox{e}^{-2u} \!\! 
\sum_{j,k,l,m} \!
\frac{(\frac{u}{2})^{j+k+l+m}}{j!\,k!\,l!\,m!} 
\mbox{e}^{-i\omega t(j-k)(l-m)}.
\label{SlinC}
\end{equation}
The sum above  runs from 0 to $\infty$ for all indexes. Since the entropy in Eq.~(\ref{SlinC}) is periodic in time, with period $T=2\pi/\omega$, we use the Fourier series and arrive at
\begin{equation}
S_L^{osc}(t) = c_0(u)
- \sum_{n=1}^\infty \;  c_n(u) \cos(n\omega t) ,
\label{FSC}
\end{equation}
where the coefficients $c_n(u)$ are given by
\begin{equation}
c_{n}(u) = 4 \, \mbox{e}^{-2u}
\sum_{k,m} \sum_{j>k} 
\sum_{l>m}\frac{(\frac{u}{2})^{j+k+l+m}}{j!\,k!\,l!\,m!} \delta_{(j-k)(l-m)}^n,
\label{Cn}
\end{equation}
representing the amplitude of the mode with frequency~$n\omega$. In Eq.~(\ref{Cn}), the Kronecker delta function, $\delta^b_a$ for integers $a$ and $b$, prompts the analysis of how $n$ decomposes as a product of two integers. We introduce the set $\Lambda_n$ composed of all distinct positive divisors of $n$ and note that the non-null terms in Eq.~(\ref{Cn}) are those for which $(j-k)$ and $(l-m)$ belong to $\Lambda_n$ and their product equals $n$. Therefore,
\begin{equation}
c_n (u) = 4 \, \mbox{e}^{-2u} \,
\sum_{\mu\in\Lambda_n} \; I_\mu(u) \, I_{\frac{n}{\mu}}(u),
\label{genesis}
\end{equation}
where $I_{\chi}(w) \equiv \sum_{k=0}^\infty [k!(\chi+k)!]^{-1} (\frac{w}{2})^{2k+\chi}$ is the modified Bessel function of the first kind. As we show next,  $c_n(u)$ is very sensitive to the primality of $n$. 

%%%%%%%%%%%%%%%%%%%%%%%%%%%%%%%%%%%%%%%%%%%%%
{\em Identifying primes.}---To show that the entanglement dynamics quantified by $S_L^{osc}(t)$ is a prime identifier, we first assume that $n$ is prime, so that $\Lambda_n=\{1,n\}$ and Eq.~(\ref{genesis}) yields $c_n (u) = c_n^{\scriptscriptstyle{(p)}} (u) $, where
\begin{equation}
c_n^{\scriptscriptstyle{(p)}} (u) \equiv 
8 \,\mbox{e}^{-2u} I_1(u) \, I_{n}(u).
\label{prime}
\end{equation}
In contrast, when $n$ is as a composite number, $\Lambda_n$ necessarily consists of 1, $n$, and, at least, one more integer. By defining the set $\Lambda'_n = \Lambda_n - \{1,n\}$, Eq.~(\ref{genesis}) becomes 
\begin{equation}
c_n(u) =  c_n^{\scriptscriptstyle{(p)}} (u) + 4 \,\mbox{e}^{-2u}
\sum_{\mu\in\Lambda'_n} I_\mu(u) \, I_{\frac{n}{\mu}}(u) \ge
c_n^{\scriptscriptstyle{(p)}} (u),
\label{composite}
\end{equation}
which holds only for $n>1$.

Inequality~(\ref{composite}) is a main result of this work. It shows that the coefficients $c_n(u)$ coincides with $c_n^{\scriptscriptstyle{(p)}}(u)$ in Eq.~(\ref{prime}) if, and only if, $n$ is prime, while for composite numbers, the amplitudes are strictly lower bounded by $c_n^{\scriptscriptstyle{(p)}} (u)$. This means that if we have a way to determine the values of $c_n(u)$, other than through the construction of $\Lambda_n$,  the application of Eq.~(\ref{composite}) can reveal the primality of $n$. Indeed, $c_n(u)$ can be evaluated theoretically via Eq.~(\ref{SlinC}) and possibly experimentally.

\begin{figure}
\vspace{-0.5cm}
\centerline{\includegraphics[width=8.5cm]{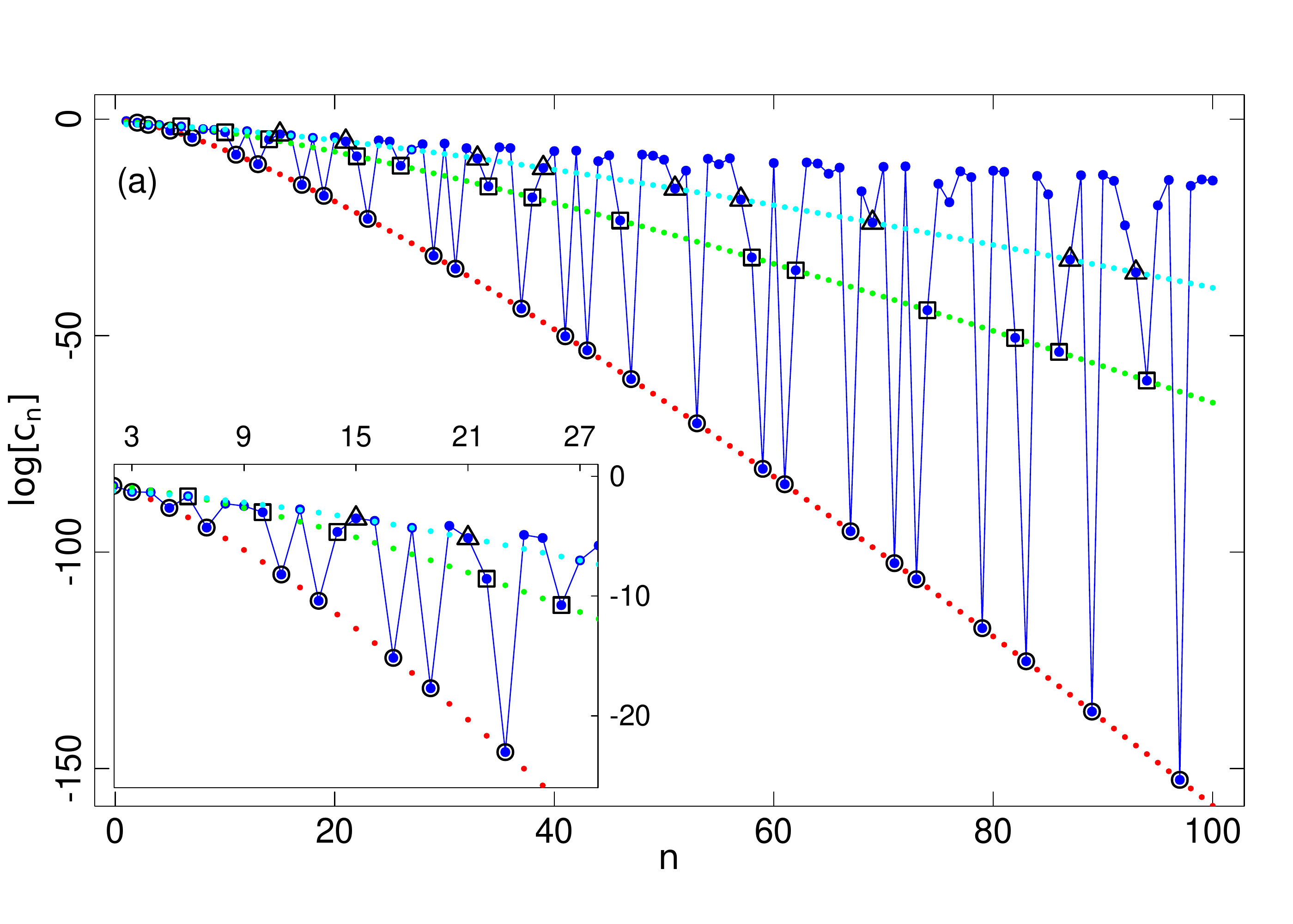}}
\vspace{-0.3cm}
\centerline{\includegraphics[width=8.5cm]{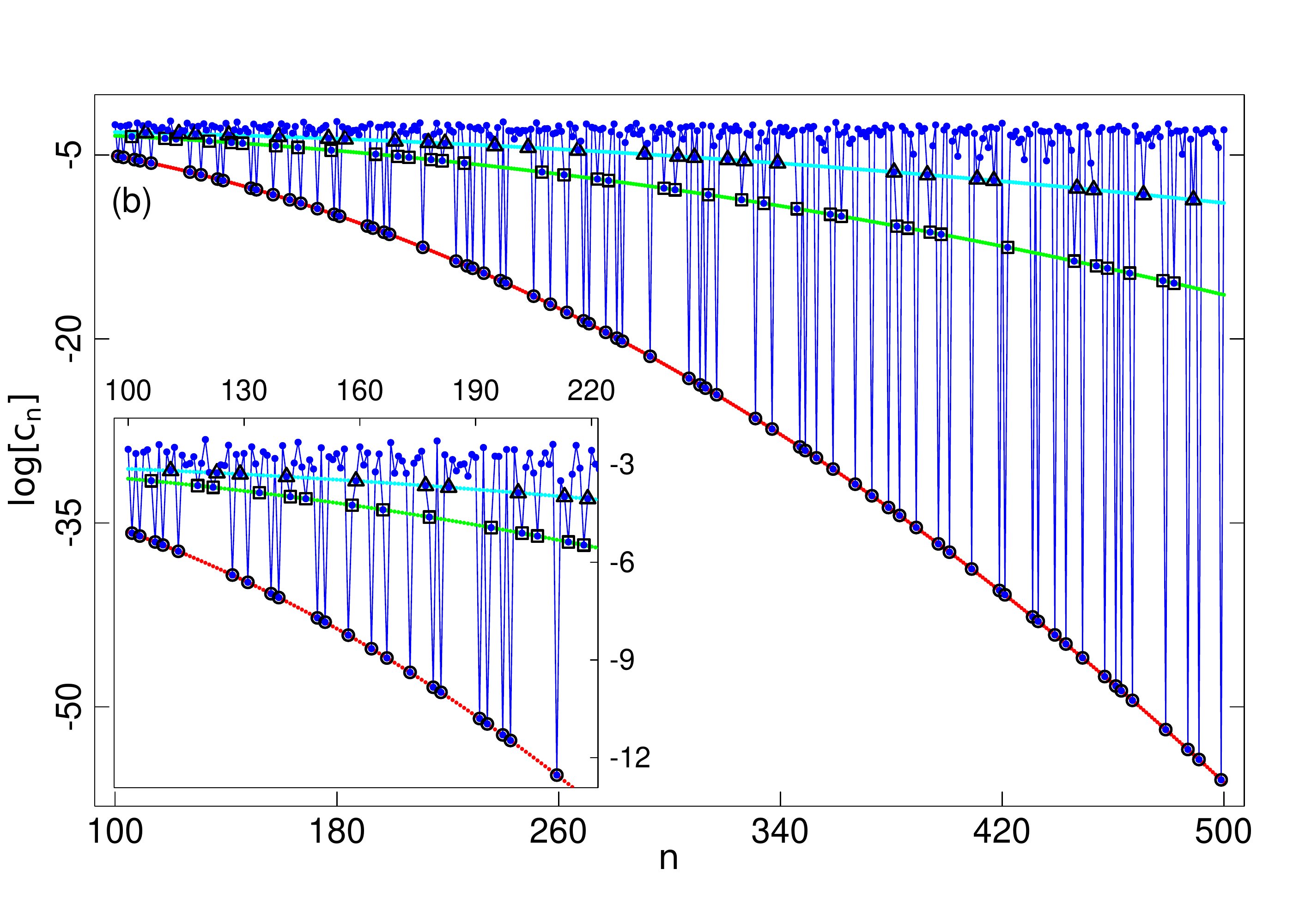}}
\vspace{-0.2cm}
\caption{Logarithm of the coefficients $c_n(u)$ as a function of $n$ for (a) $u=1$ and (b) $u=10^3$. The blue dots, connected with a blue solid line to guide the eye, represent $c_n(u)$, and the red points indicate the lower bound $c_n^{\scriptscriptstyle{(p)}}(u)$. For each prime $n$, the blue dots are encircled with black circles and touch the red dotted line. Green and cyan lines are for $c_n^{\scriptscriptstyle{(f_2)}}(u)$ and $c_n^{\scriptscriptstyle{(f_3)}}(u)$, respectively. Black squares [triangles] enclose the coefficients of the family $f_2$ [$f_3$]. Inset plots show a magnified region of the respective main graph. }
\label{fig1}
\end{figure}

In Fig.~\ref{fig1}, we mark the values of $c_n(u)$ with blue dots. For each $n$ identified as a prime, the dots are encircled with a black circle. All values of $c_n(u)$ corresponding to primes coincide with the red dotted line that represents the lower bound $c_n^{\scriptscriptstyle{(p)}} (u)$ and can, therefore, be clearly distinguished. In Fig.~\ref{fig1}~(a), we show results for $u=1$, but  $c_n(1)$ becomes too small when $n$ is large, so in Fig.~\ref{fig1}~(b), we use $u=10^3$ and larger values of $n$.

%%%%%%%%%%%%%%%%%%%%%%%%%%%%%%%%%%%%%%%%%%%%%
{\em Semiprimes.}--- Figure~\ref{fig1} also reveals the square-free semiprimes, including the integer 2, which we denote by family $f_2$. For these numbers,  $\Lambda_n=\{1,2,P_n,n\}$, where $P_n=n/2\neq 2$ is a prime, and $c_n(u)=c_n^{\scriptscriptstyle{(f_2)}} (u)$, where
\begin{equation}
c_n^{\scriptscriptstyle{(f_2)}} (u) \equiv 
8 \,\mbox{e}^{-2u} \left[I_1(u) \, I_{n}(u) + I_2(u) \, I_{\frac{n}{2}}(u)\right].
\label{semi2}
\end{equation}
For a composite $n$ that presents the divisors: 1, 2, $n/2$, $n$, and at least one more integer, 
\begin{equation}
c_n^{\scriptscriptstyle{(2)}} (u) \equiv  c_n^{\scriptscriptstyle{(f_2)}} (u) +
4\,\mbox{e}^{-2u} 
{\sum_{\mu\in\Lambda^{'\scriptscriptstyle{(2)}}_n}} 
I_\mu(u) \, I_{\frac{n}{\mu}}(u) > c_n^{\scriptscriptstyle{(f_2)}} (u) ,
\nonumber
%\label{n2}
\end{equation}
where $\Lambda^{'\scriptscriptstyle{(2)}}_n = \Lambda_n - \{ 1, 2, n/2, n \}$. The semiprimes of family $f_2$ are identified with black squares in Fig.~\ref{fig1} and they all fall on the curve $c_n^{\scriptscriptstyle{(f_2)}} (u) $, indicated with a green line. This line consists of a lower bound for any integer composed by 2, except for $2$, $2^2$, and $2^3$.

It is straightforward to extend the analysis above to any other family of semiprimes. The particular case of the square-free semiprimes containing the number 3, denoted by $f_3$, is shown in Fig.~\ref{fig1}. The components of family $f_3$ are marked with black triangles and they are exactly located over the cyan curve $c_n^{\scriptscriptstyle{(f_3)}} (u)$~\cite{spf3}.

%%%%%%%%%%%%%%%%%%%%%%%%%%%%%%%%%%%%%%%%%%%%%
{\em Interacting spins.---} Analogously to the system of coupled harmonic oscillators, we now show that two interacting spins with a large quantum number $S$ is another physical system that can be used to identify primes. In the  Hamiltonian of Eq.~(\ref{Hoperator}), we assume that $H_A = \hbar \omega_0 S_A^{z}$, where $S_A^{z}$ is the $z$-component of the spin operator of part $A$ and $S_A^{z}|m_{A}\rangle=  m_{A} |m_{A}\rangle$, and equivalently for part $B$. The initial state is the product of spin coherent states~\cite{gazeau}, $|\Psi(0) \rangle =  |s_{A}\, s_{B} \rangle$, where
\begin{equation}
\nonumber
|s_{A}\, s_{B} \rangle 
\equiv N_s \!\!\!\!\!\!
\sum_{n_A,n_B=0}^{2S} \!\! \frac{s_{A}^{n_A} s_{B}^{n_B}}{n_{A}!n_{B}!} 
\sqrt{\textstyle\binom{2S}{n_{A}} \binom{2S}{n_{B}}}
|n_{A}-S,n_{B}-S\rangle,
\end{equation}
where $N_s=(1+u^2)^{-2S} $ and we chose $|s_{A}|^2= |s_{B}|^2 = u$.

The evolving linear entanglement entropy~(\ref{Slin}) becomes
\begin{equation}
S_L^{spin}(t) = 1 - \! \! \! \!
\! \sum_{j,k,l,m=0}^{2S} \! \! \! \! \xi_{j,k,l,m} \,
u^{j+k+l+m} \mbox{e}^{-i\omega t(j-k)(l-m)},
\label{Sspin}
\end{equation}
where $\xi_{j,k,l,m}\equiv \,
{\textstyle \binom{2S}{j} \binom{2S}{k} \binom{2S}{l} \binom{2S}{m}} (1+u)^{-8S}$. The Fourier series of Eq.~(\ref{Sspin})  has the same structure as Eq.~(\ref{FSC}), but with the coefficients
\begin{equation}
\nonumber
\bar{c}_{n}(u) =
4 \sum_{k,m=0}^{2S-1} \sum_{j>k}^{2S} \sum_{l>m}^{2S} 
\xi_{j,k,l,m} \, u^{j+k+l+m} \delta_{(j-k)(l-m)}^n.
\end{equation}
The nonvanishing terms in the equation above are again those for which $\mu\equiv(j-k)$ and $\nu\equiv(l-m)$ belong to the set $\Lambda_n$ of divisors of $n$. However, this system is bounded, so the upper limits of the above summations add new constraints to $\mu$ and $\nu$. Given a value for $k$ and $m$, we have that $\mu\in \mathcal{I}_{k}\equiv [1,2S-k]$ and $\nu\in \mathcal{I}_{m}\equiv  [1,2S-m]$. In addition, since  $\delta_{\mu\nu}^n$  implies that $\nu=n/\mu$, we also get that $\mu\in \mathcal{I}_{m,n}\equiv[\frac{n}{2S-m}, n]$. Therefore,
\begin{equation}
\bar{c}_{n}(u) = 4\sum_{k,m=0}^{2S-1} \, \sum_{\mu\in\bar{\Lambda}^{\scriptscriptstyle{(k,m)}}_n} 
\!\!\! \bar{\xi} \, u^{2k+2m+\mu+n/\mu} ,
\label{cns}
\end{equation}
where $\bar{\xi} \equiv  \xi_{k+\mu,k,m+\frac{n}{\mu},m}$ and $\bar{\Lambda}_n^{\scriptscriptstyle{(k,m)}}$ is the set of divisors of $n$ that satisfies all constraints commented above, 
\begin{equation}
\bar{\Lambda}_n^{\scriptscriptstyle{(k,m)}} \equiv \Lambda_n \cap \mathcal{I}_{k} \cap \mathcal{I}_{m,n}.
\label{Lmk}
\end{equation}

Contrary to $S_L^{osc}(t)$ in Eq.~(\ref{FSC}), there is a finite number of Fourier modes in $S_L^{spin}(t)$ due to the finite Hilbert space for the spin system. For $n>4S^2$, we have that $\mathcal{I}_{k} \cap \mathcal{I}_{m,n} =\emptyset$ for any value of $k$ and $m$, so $\bar{\Lambda}_n^{\scriptscriptstyle{(k,m)}}=\emptyset$ and $\bar{c}_n(u)=0$. In what follows, to assess the primality of $n$, we conveniently divide the interval $1<n\le 4S^2$ in two regions of interest, region I ($1 < n \le 2S$) and region II ($2S < n \le 4S$). In the complementary range $4S < n \le 4S^2$, one cannot find a behavior that distinguishes primes from composite numbers. There, even though all amplitudes of the Fourier modes for a prime $n$ vanish, the same can also happen to some composite numbers.

Region I ($1< n \le 2S$): A careful inspection of Eq.~(\ref{Lmk}) reveals that $1\in\bar{\Lambda}^{(k,m)}_{n}$ for all values of $k$ and $0\le m\le 2S-n$, while $n\in\bar{\Lambda}^{(k,m)}_{n}$ for all values of $m$ and $0\le k\le 2S-n$. Applying these results to Eq.~(\ref{cns}), we find that, for $n$ prime, $\bar{c}_n(u)=\bar{c}^{{\scriptscriptstyle (p)}}_{n,I}(u)$, where
\begin{equation}
\bar{c}_{n,I}^{{\scriptscriptstyle (p)}}(u) \equiv 8\,(1+u)^{-8S} G_1(u) G_n (u),  
\label{cnsp}
\end{equation}
with $G_\chi(w) \equiv \sum_{k=0}^{2S-\chi} \binom{2S}{k}\binom{2S}{k+\chi} w^{2k+\chi}$. On the other hand, for a composite number $n>1$ inside region I, we have that $\bar{c}_n(u) = \bar{c}_{n}^{{\scriptscriptstyle (p)}}(u)+\bar{d}_{n,I}(u)$, where
\begin{equation}
\bar{d}_{n,I}(u) \equiv 4
{\sum_{k,m=0}^{2S-1} \, \sum_{\mu\in{\bar{\Lambda}^{\prime\scriptscriptstyle{(k,m)}}_n}}}
\, \bar{\xi} \, u^{2k+2m+\mu+n/\mu}   
\label{cnsc}
\end{equation}
and ${\bar{\Lambda}^{\prime\scriptscriptstyle{(k,m)}}_n} \equiv \Lambda'_n \cap \mathcal{I}_{k} \cap \mathcal{I}_{m,n}$, where $\Lambda'_n$ was defined above Eq.~(\ref{composite}). One can show that at least for $k=m=0$, there is a positive non-null term in Eq.~(\ref{cnsc}). Therefore, for a composite number, we necessarily have $\bar{c}_n(u) > \bar{c}_{n,I}^{{\scriptscriptstyle (p)}}(u)$ and this inequality can be employed in the search for prime numbers in region I.

Region II ($2S<n\le 4S$): In this case, prime modes are not included in the Fourier series of Eq.~(\ref{Sspin}). Indeed, for a prime number $n$, we now have that $1\not\in \mathcal{I}_{m,n}$ and $n\not\in \mathcal{I}_{k} $, which implies that $\bar{\Lambda}_n^{\scriptscriptstyle{(k,m)}}=\emptyset$ and $\bar{c}_n(u)=0$.  For a composite $n$ in the same interval, similarly to region I, there is at least one integer in $\Lambda_n$, in addition to 1 and $n$, so that $\bar{c}_n(u) > 0$. So we can distinguish primes from composite numbers by checking if $\bar{c}_n(u)=0$ or $\bar{c}_n(u) \neq 0$.

\begin{figure}
    \centering
    \includegraphics[width=8.5cm]{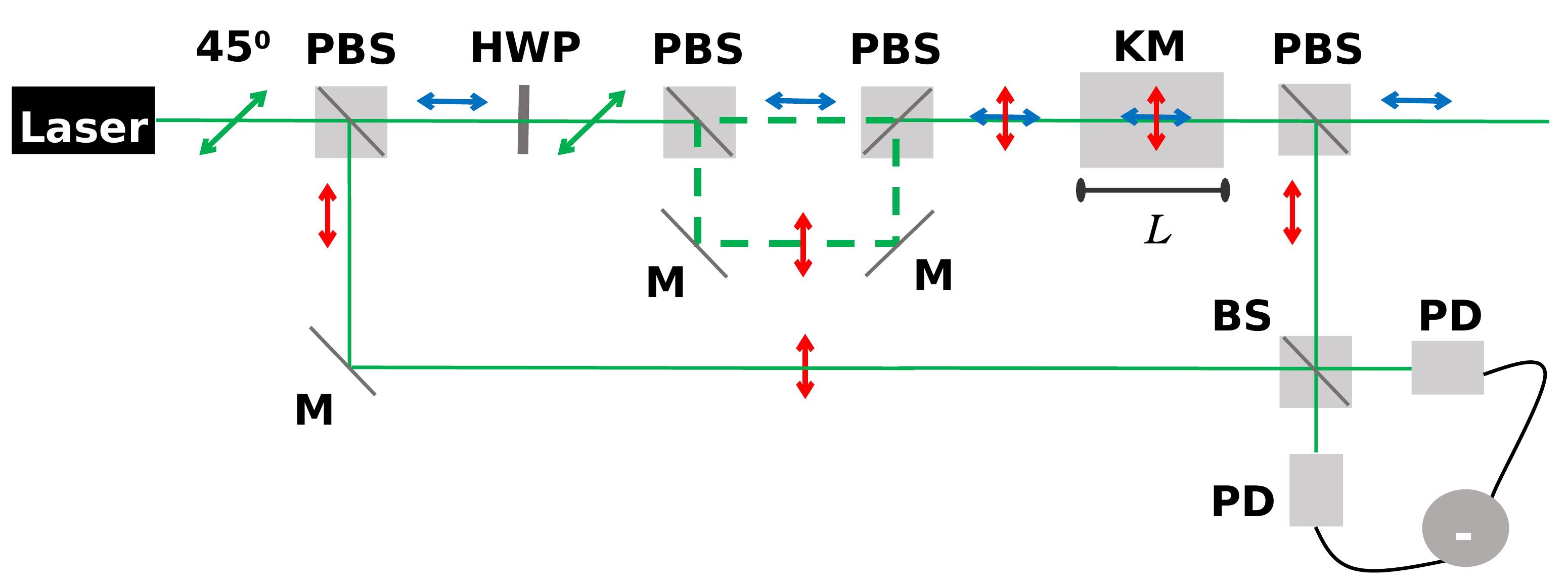}
    \caption{Setup for the identification of prime numbers. The symbols BS, HWP, PBS, M, KM, and PD refer to beam splitter, half-wave plate, polarizing beam splitter, mirror, Kerr medium, and photodiode, respectively. See text for details.}
    \label{fig_setup}
\end{figure}

%%%%%%%%%%%%%%%%%%%%%%%%%%%%%%%%%%%%%%%%%%%%%
{\em Experimental proposal.---} The Hamiltonian for the two coupled harmonic oscillators studied above also describes the interaction between two optical fields via a Kerr nonlinear medium~\cite{scully, yamamoto}. In Fig.~\ref{fig_setup}, we show the sketch of an experimental setup for implementing this optical system. A laser beam of frequency $\omega_0$ and linear polarization at $45^\circ$ is sent to a polarizing beam splitter (PBS). The vertically polarized~($\updownarrow$) component goes directly to the homodyne detection scheme~\cite{HomodyneDetect}, shown in the lower right corner of the figure, to act as the local field. The horizontally polarized ($\leftrightarrow$) component passes through a half waveplate (HWP), which rotates the polarization to $45^\circ$, and enters an unbalanced Mach-Zehnder interferometer, identified in Fig.~\ref{fig_setup} with dashed lines. In the interferometer, after the PBS, the $\updownarrow$~component beam propagates through the short path, and the $\leftrightarrow$~one goes through the long course. The path difference is longer than the coherence length, so that the recombined $\leftrightarrow$ and $\updownarrow$ beams at the interferometer output PBS are separable and no longer result in a pure mode with linear diagonal polarization. Next, these two beams are injected in the nonlinear Kerr medium (KM) of length $L$ and Kerr optical nonlinearity~$\chi^{(3)}$. During the propagation time inside the KM, each beam will experience a modified index of refraction due to the action of the other beam. This is how the coupling between the oscillators is physically implemented. By varying the length $L$, one changes the interaction time $t$. After passing KM, the $\leftrightarrow$ and $\updownarrow$ beams are split again. To identify the prime numbers, we measure one of the beams and ignore the other, which is equivalent to performing the trace over one of the interacting systems. We can analyze any of the two beams because of the interaction symmetry. In Fig.~\ref{fig_setup}, we choose the $\updownarrow$ beam, which goes to the homodyne detector, where quantum state tomography is performed to reconstruct the density matrix $\rho_A$. With the reduced density matrix, we calculate the linear entanglement entropy. After several experimental realizations spanning the needed values of~$t$, curves like those in Fig.~\ref{fig1} can be generated.

%%%%%%%%%%%%%%%%%%%%%%%%%%%%%%%%%%%%%%%%%%%%%
{\em Conclusion.---} This paper shows that by analyzing the bipartite entanglement in time, one can identify prime and semiprime numbers in $\mathbb{N}$. The main ingredient is a Hamiltonian composed of two parts, $A$ and $B$, with equidistant energy levels for $H_A$ and $H_B$. We discussed how this idea could be implemented taking advantage of  an existing experimental setup.

Our work resonates with the Hilbert-P\'olya conjecture in the sense of proposing a physically measurable quantity to determine prime numbers. We speculate that there may be a way to connect our results with $\zeta(s)$. In fact, by defining $A(n,u) \equiv c_n(u) - c^{\scriptscriptstyle(p)}_n(u)$ for a fixed~$u$, the task of counting primes along $\mathbb{N}$ is equivalent to counting the zeros of $A(n,u)$. This alternative method to build $\pi(n)$ could be used to study the zeros of $\zeta(s)$ by means, for example, of a hypothetical inversion of the Riemann series that connects $s_0$ with $\pi(n)$, shedding new light on the Riemann hypothesis.
% https://empslocal.ex.ac.uk/people/staff/mrwatkin/zeta/encoding1.htm

Our last remark concerns the semiclassical approaches originated from the ideas of Berry and Keating~\cite{keating99}. The semiclassical version of the linear entanglement entropy was addressed in~\cite{arlans,matheus} for the same two physical systems treated here. In those papers, entanglement is reproduced by summing over sets of classical trajectories determined by the solutions of a given transcendental equation. The purely quantum formalism presented here encourages the re-examination of those works aiming at connecting those solutions with the distribution of primes.

%--------------------------------------------------------------------
%--------------------------------------------------------------------
\section*{Acknowledgements}

This study was financed by the Brazilian agencies: Funda\c{c}\~{a}o de Amparo \`{a} Pesquisa do Estado de Santa Catarina (FAPESC -- DOI 501100005667), Coordena\c{c}\~ao de Aperfei\c{c}oamento de Pessoal de N\'{i}vel Superior (CAPES -- DOI 501100002322), Conselho Nacional de Desenvolvimento Cient\'{\i}fico e Tecnol\'ogico (CNPq -- DOI 501100003593), and Instituto Nacional de Ci\^encia e Tecnologia de Informa\c{c}\~ao Qu\^antica (INCT-IQ 465469/2014-0). LFS was supported by the NSF CCI grant (Award Number 2124511). The authors also thank Gisele T. Paula and Jo\~ao V. P. Poletto for their valuable discussion at the very beginning of the work.

%--------------------------------------------------------------------
%--------------------------------------------------------------------


\begin{thebibliography}{99}
\bibitem{path}
The path of integration $\gamma$, defined in the $x$-complex plane, starts at $x\to\infty+i\epsilon$, with $\epsilon\to0_+$, goes to the origin parallelly to the real axis, encircles it in the counter clock sense, and returns to the final point $x\to\infty-i\epsilon$, also parallelly to the real axis.
\end{thebibliography}
\end{document}